\documentclass[fleqn,usenatbib]{mnras}

\usepackage{newtxtext,newtxmath}

\usepackage[T1]{fontenc}

\DeclareRobustCommand{\VAN}[3]{#2}
\let\VANthebibliography\thebibliography
\def\thebibliography{\DeclareRobustCommand{\VAN}[3]{##3}\VANthebibliography}

\usepackage{graphicx}	
\usepackage{amsmath}	

\usepackage{amssymb}	
\usepackage{booktabs}

\title[Possible GeV gamma-ray emission from CTA 1]{Possible GeV gamma-ray emission from the pulsar wind nebula in CTA 1}

\author[Zhou et al.]{
Liancheng Zhou,$^{1}$
Keyao Wu,$^{1}$
Yunlu Gong,$^{1}$
and Jun Fang$^{1}$\thanks{E-mail: fangjun@ynu.edu.cn }
\\
$^{1}$Department of Astronomy, Yunnan University, Kunming 650091, China;
}

\date{Accepted 2024 March 6. Received 2024 March 3; in original form 2023 November 6}

\pubyear{2015}

\begin{document}
\label{firstpage}
\pagerange{\pageref{firstpage}--\pageref{lastpage}}
\maketitle

\begin{abstract}
We report a detection of GeV $\gamma$-ray emission potentially originating from the pulsar wind nebula in CTA 1 by analyzing about 15 yr of Fermi Large Area Telescope data. By selecting an energy range from 50 GeV to 1 TeV to remove contamination from the $\gamma$-ray pulsar PSR J0007+7303, we have discovered an extended $\gamma$-ray source with a TS value of $\sim$ 44.94 in the region of CTA 1. The obtained flux is measured to be 6.71 $\pm$ 2.60 $\times$ $10^{-12}$ erg $\mathrm{cm}^{-2}$ $\mathrm{s}^{-1}$ with a spectral index of 1.61 $\pm$ 0.36, which allows for a smooth connection with the flux in the TeV band. CTA 1 is also considered to be associated with 1LHAASO J0007+7303u, which is an Ultra-High-Energy source listed in the recently published catalog of the Large High Altitude Air Shower Observatory. We assume that the radiation originates from the pulsar wind nebula and that its multi-wavelength spectral energy distribution can be explained well with a time-dependent one-zone model.
\end{abstract}




\section{Introduction} \label{sec:Introduction}

The detection and characterization of Ultra-High-Energy (UHE; $>$ 0.1 PeV) $\gamma$-ray sources have significant implications for our understanding of the origin and acceleration mechanisms of cosmic rays. The Large High Altitude Air Shower Observatory (LHAASO) is a revolutionary project that aims to explore the high-energy universe with remarkable sensitivity \citep{2019arXiv190502773C}. The LHAASO collaboration has made a breakthrough by reporting the detection of 12 UHE $\gamma$-ray sources with statistical significance exceeding 7$\sigma$ and a maximum energy of up to 1.4 PeV \citep{2021Natur.594...33C}, sparking further exploration of UHE gamma-ray sources. Recently, the first catalog of $\gamma$-ray sources detected by LHAASO was published \citep{2023arXiv230517030C}, containing a total of 90 $\gamma$-ray sources with an extended size smaller than 2$^\circ$ and a significance of detection at a $>$ 5$\sigma$ significance level. Among these sources, 43 exhibit UHE emission at a $>$ 4$\sigma$ significance level. 1LHAASO J0007+7303u is a TeV source in the first LHAASO catalog, detected by both the Water Cherenkov Detector Array (WCDA) and the 1 km$^2$ array (KM2A) as a UHE source with a significance of TS$_{100}$ = 171.6. This source is considered to be associated with the shell-type supernova remnant (SNR) CTA 1 (approximately 0.$^\circ$12 away) \citep[][]{2023arXiv230517030C}.

CTA 1 (G119.5+10.2) was first proposed as a composite SNR through radio observations by \cite{1960PASP...72..237H}, and it was first detected in the X-ray band by \cite{1995ApJ...453..284S}. In the center of the region, there was a faint compact point source called RX J0007.0+7303, which was considered to be a pulsar candidate powering the nebula \citep{1995ApJ...453..284S,1997ApJ...485..221S}. Deeper observations of RX J0007.0+7303 were conducted using $XMM-Newton$, ASCA satellites \citep{2004ApJ...601.1045S}, and the $Chandra$ observatory \citep{2004ApJ...612..398H}, revealing that it is a point source embedded in a nebula with a jet-like feature. Additionally, the source 3EG J0010+7309, which is associated with RX J0007.0+7303, has been reported as a radio-quiet $\gamma$-ray pulsar candidate \citep{1996A&AS..120C..95M}. Although no $\gamma$-ray pulsations had been found, the constant $\gamma$-ray flux, the hard power-law spectrum, and the steepening of the spectrum above approximately 2 GeV, confirmed that the source was a $\gamma$-ray pulsar \citep{1998MNRAS.295..819B}. The Fermi Large Area Telescope (Fermi-LAT) satellite then discovered a $\gamma$-ray pulsar (PSR J0007+7303) in CTA 1, which had a period of 316.86 ms and a period derivative of 3.614 $\times 10^{-13}$ s s$^{-1}$, making it the first pulsar discovered by $\gamma$-rays \citep{2008Sci...322.1218A}. Subsequently, X-ray pulsations were detected \citep{2010ApJ...725L...6C, 2010ApJ...725L...1L}. However, neither radio nor optical counterparts of RX J0007.0+7303 were detected \citep{2013MNRAS.430.1354M}.

The TeV $\gamma$-ray emission from CTA 1 was discovered by \cite{2013ApJ...764...38A} using the $VERITAS$ observatory. The extended source, VER J0006+729, was detected at a 6.5$\sigma$ significance level, with a two-dimensional Gaussian of 0.30$^\circ$ $\times$ 0.24$^\circ$ at a distance of 5$'$ from PSR J0007+7303. Previously, \cite{2012ApJ...744..146A} reported potentially GeV extended emission after analyzing 2 yr of Fermi-LAT data in the off-pulse emission of PSR J0007.0+7303, but they only provided upper limits. On the other hand, \cite{2016ApJ...831...19L} using 7 yr of Fermi-LAT data and updated response functions, did not detect GeV emission compatible with CTA 1 or the highly energetic ($>$ 100 GeV) pulsar wind nebula. Instead, they provided an upper limit. In a more recent study, \cite{2018ApJS..237...32A} reported GeV emission with a radius of approximately 0.98$^\circ$, spatially consistent with the radio emission from CTA 1. However, there was a discrepancy in the flux normalization between the GeV and TeV spectra. In this work, we reanalyze the GeV emission from CTA 1 using about 15 yr of Fermi-LAT data.

First, Section \ref{sec:Introduction} serves as the introduction. Then, in Section \ref{sec:FERMI-LAT Data Analysis}, we present a detailed analysis of the Fermi-LAT data. Next, Section \ref{Discussion} contains specific discussions related to the analysis. Lastly, Section \ref{Summary} offers a summary of the paper.

\section{FERMI-LAT Data Analysis} \label{sec:FERMI-LAT Data Analysis}
\subsection{Data Selection}
\label{Data Selection}
We analyzed about 15 yr (from 2008 August 4 to 2023 May 24) of data toward CTA 1 in a 10$^\circ$ radius centered on PSR J0007+7303 \citep[logged as 4FGL J0007.0+7303 in 4FGL-DR3;][]{2022ApJS..260...53A} with the available software \texttt{Fermipy} \citep{2017ICRC...35..824W}. The newest response functions \texttt{P8R3$\_$SOURCE$\_$V3} and the Pass 8 events with “source” event class (evtype = 3 and evclass = 128) was selected, while the expression of “(DATA$\_$QUAL $>$ 0)$\&\&$(LAT$\_$CONFIG==1)” was used to filter a good time interval. The maximum zenith angle was fixed to $90^\circ$ to limit contamination of $\gamma$-rays from the Earth’s limb. To search for potential $\gamma$-rays from the PWN, we chose an energy band of 50 GeV$-$1 TeV to suppress the contamination of PSR J0007+7303.

The sources in the Data Release 3 of the fourth Fermi-LAT source catalog \citep[4FGL-DR3;][]{2022ApJS..260...53A} were considered in the analysis. The isotropic $\gamma$-ray background emission (\texttt{iso$\_$P8R3$\_$SOURCE$\_$V3$\_$v1}) and the diffuse Galactic interstellar emission (\texttt{gll$\_$iem$\_$v07}) were used to construct the background model. In the analysis that follows, the normalization and spectral parameters are free for the sources within $3^\circ$ in the background model. The normalization parameters of the isotropic and Galactic components are also left free.

\subsection{Spatial Analysis}
\label{Spatial Analysis}

\begin{figure*}
\includegraphics[width=1.6\columnwidth]{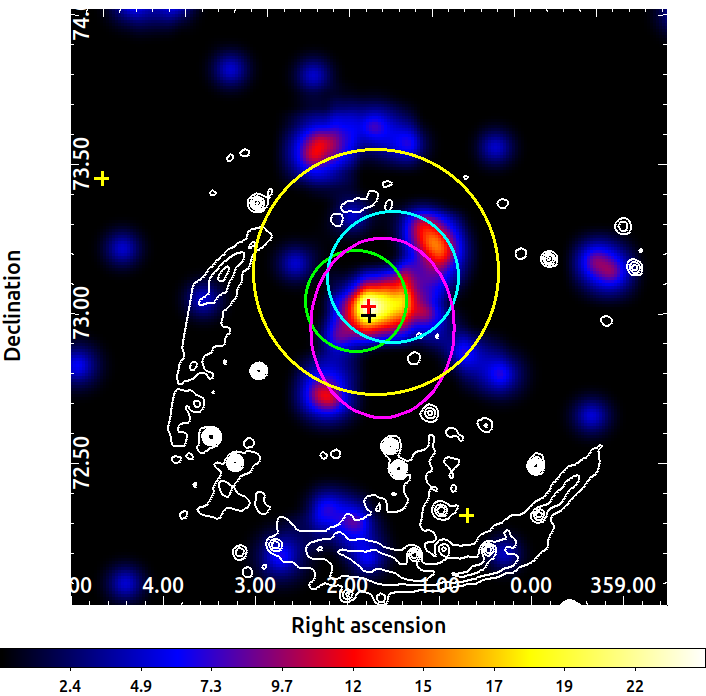}
    \caption{The excess TS map in the 50 GeV$-$1 TeV energy band centred on 4FGL J0007.0+7303 with each pixel of 0.$^\circ$01. All the background sources are removed in the map. The red, black and yellow crosses indicate the position of 4FGL J0007.0+7303, the best-fit position of the point-source model from this work and other 4FGL sources, respectively. The cyan circle (a radius of $\sim$ 0$^\circ$.17) shows the 95\% statistic upper limits of the extension detected by WCDA, and the green circle (a radius of $\sim$ 0$^\circ$.22) shows the 39\% containment radius of the 2D-Gaussian model detected by KM2A \citep{2023arXiv230517030C}. Magenta ellipse of 0.$^\circ$30 $\times$ 0.$^\circ$24 indicates the VERITAS's observation \citep{2013ApJ...764...38A}. The yellow circle indicates the best-fit extension radius of the GeV excess under the uniform disk model. The white contours correspond to the radio observation of the SNR shell \citep{1997A&A...324.1152P}. The map was smoothed using a Gaussian function with a kernel radius of 0.$^\circ$06.}
    \label{fig:tsmap}
\end{figure*}

\begin{figure*}
    \centering

    \begin{minipage}{0.49\linewidth}
     \centering
     \includegraphics[width=0.9\linewidth]{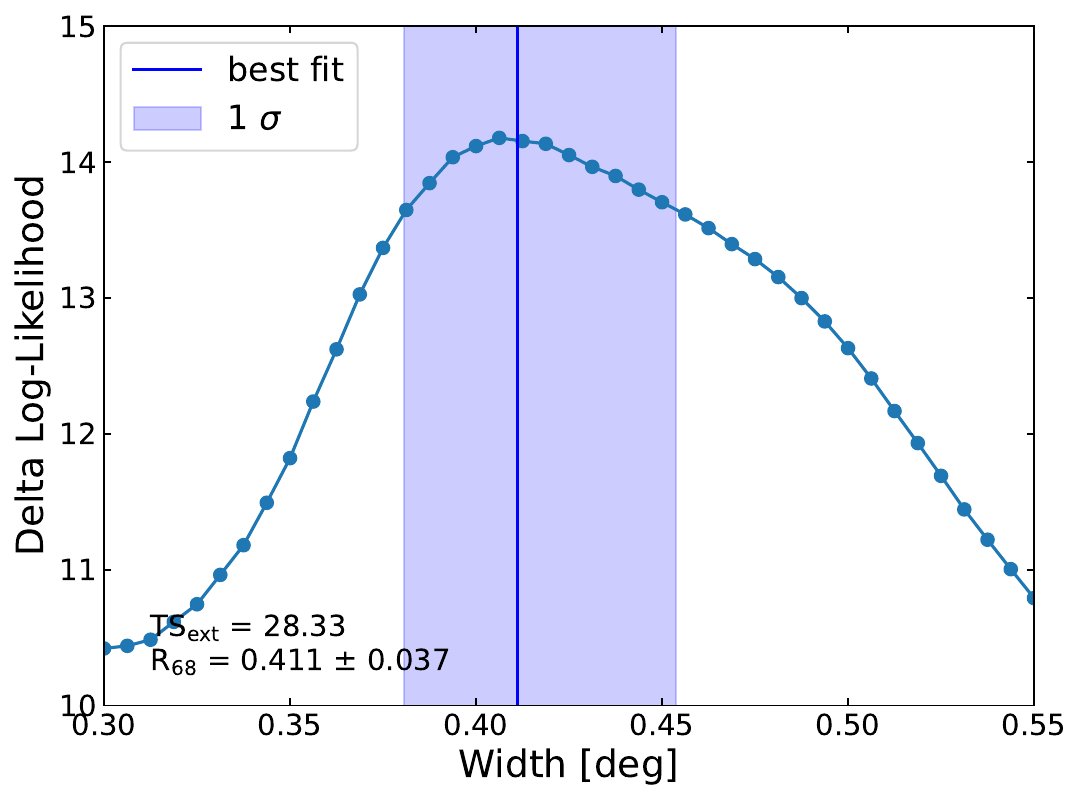}
    \end{minipage}
    \begin{minipage}{0.49\linewidth}
     \centering
     \includegraphics[width=0.9\linewidth]{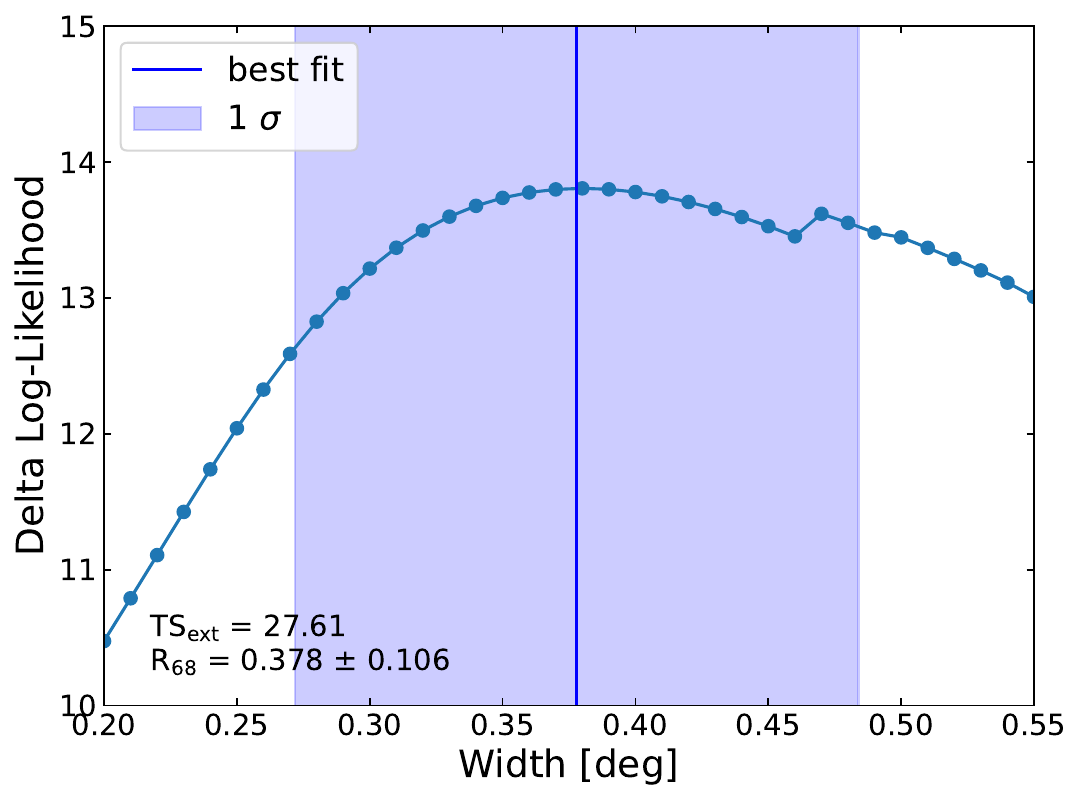}
    \end{minipage}

    \caption{Extension likelihood curves for uniform disk model (left), and 2D Gaussian (right). The best-fit for the extension is represented by the blue vertical line, and the 1$\sigma$ uncertainty is shown as a purple band.}

    \label{fig:extension analysis}
\end{figure*}

To investigate the possibility of detecting the radiation originating from PWN in CTA 1, we first generated a test statistic (TS) map centred on 4FGL J0007.0+7303 \citep[R.A. = 1.$^\circ$768, Decl. = 73.$^\circ$051;][]{2022ApJS..260...53A}. In order to minimize contamination from PSR J0007+7303, we set an energy threshold of 50 GeV. We performed a fitting procedure for the background sources and subtracted 4FGL J0007.0+7303 to create the excess TS map in the energy range of 50 GeV to 1 TeV. Here, 4FGL J0007.0+7303, which corresponds to PSR J0007+7303, is a point source listed in the 4FGL catalog with a PLSuperExpCutoff4 spectrum \citep{2022ApJS..260...53A}.

In Figure \ref{fig:tsmap}, a significant excess of GeV $\gamma$-ray radiation is evident at the observed positions of the other detectors. Based on LHAASO's observations \citep{2023arXiv230517030C}, 1LHAASO J0007+7303u was detected by KM2A as an extended source located at R.A. = 1$^\circ$.91 and Decl. = 73$^\circ$.07 with $r_{\mathrm{39}}$ $\sim$ 0$^\circ$.17 and by WCDA as a point source located at R.A. = 1$^\circ$.48 and Decl. = 73$^\circ$.15 with a 95\% error radius of $\sim$ 0$^\circ$.22. The locations can be seen in Figure \ref{fig:tsmap}, denoted by cyan and green circles. Meanwhile, magenta ellipse of 0.$^\circ$30 $\times$ 0.$^\circ$24 indicates the VERITAS's observation \citep{2013ApJ...764...38A}. To determine the best morphological model describing the excess radiation, we utilized $gtfindsrc$ to calculate the best-fit location based on the initial position with the maximum TS value. The best-fit location for the excess radiation was R.A. = 1$^\circ$.76 and Decl. = 73$^\circ$.02, with a 1$\sigma$ error of 0$^\circ$.06, marked in Figure \ref{fig:tsmap} as a black cross and labelled as SrcA for convenience.

We first tested the power-law spectrum point-source model, with a spectral index of 2.0, at the best-fit position. Spatial extension models such as the uniform disk and the 2D Gaussian models were also applied to judge the best-fit model. We applied both models to the extension analysis of SrcA using the $gta.extension$ programme in $Fermipy$ \citep{2017ICRC...35..824W}. As a result, SrcA obtained a $TS_{\mathrm{ext}}$ value of 27.61 in the 2D Gaussian template and 28.33 in the uniform disk template. Here, the $TS_{\mathrm{ext}}$ value is defined as $TS_{\mathrm{ext}}$ = 2 $(\ln\mathcal{L}_\mathrm{ext}-\ln\mathcal{L}_\mathrm{ps})$, where the $\ln\mathcal{L}_\mathrm{ext}$ and $\ln\mathcal{L}_\mathrm{ps}$ represent the maximum likelihood values of the extended and point-like model, respectively \citep[][]{2012ApJ...756....5L}. Therefore, we will chose the extended model to describe SrcA as $TS_{\mathrm{ext}}$ $\geqslant$ 16 (a significance level of 4 $\sigma$).

We calculated the delta log-likelihood value, defined as $\ln\mathcal{L}_\mathrm{ext}-\ln\mathcal{L}_\mathrm{ps}$, for various radii in order to determine the optimal extended model. The distribution of delta log-likelihood values for the various templates is shown in Figure \ref{fig:extension analysis}. Comparing the two templates, we found that the disk template had a higher $TS_{\mathrm{ext}}$ value and less uncertainty. As a result, we selected the disk template to analyze the excess $\gamma$-ray radiation in this region. Using the best-fit positions and extension of the different extended models, we performed a likelihood analysis again. During the fitting process, the spectrum type of power-law $dN/dE=N_0\left( E/E_0 \right) ^{-\varGamma}$ was used. The results of the fitting, which including the point-source model's, are conveniently presented in Table \ref{tab:spatial distribution analysis}.

\subsection{Spectral Analysis}
\label{Spectral Analysis}

\begin{figure*}
\includegraphics[width=1.4\columnwidth]{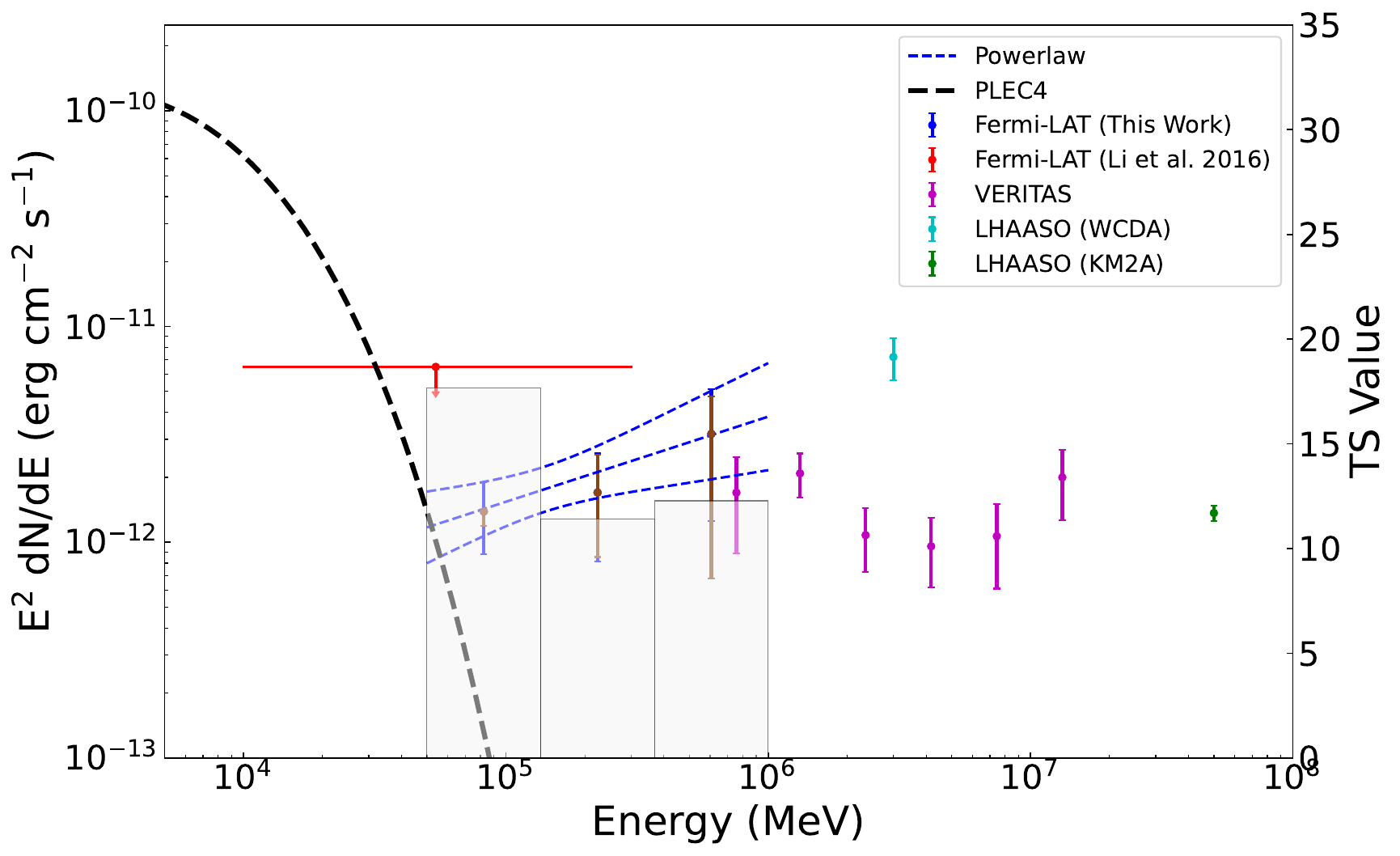}
    \caption{The $\gamma$-ray SED of CTA 1. The black dashed line indicates the energy spectrum of the PSR J0007+7303 described by a PLEC4 model. The blue dots with 1$\sigma$ statistic error and the systematic error in brown represents the results of the Fermi-LAT data analysis in this work (between the 50 GeV$-$1 TeV energy band), while the best-fit spectrum is shown by the blue lines and the TS value in each energy bin is represented by the gray-shaded region. The red line indicates the flux upper limit in the 10$-$300 GeV energy range \citep{2016ApJ...831...19L}. The cyan and green dots with statistical uncertainties represent the results of LHAASO's observations detected by WCDA and KM2A, respectively \citep{2023arXiv230517030C}. The magenta points from $VERITAS$'s observation indicate the spectrum of VER J0006+729 \citep{2013ApJ...764...38A}.}
    \label{fig:sedHigh}
\end{figure*}

\begin{table*}
	\centering
	\caption{Spatial Distribution Analysis of SrcA Using Different Models in the 50-1000 GeV Energy Range}
	\setlength{\tabcolsep}{15pt}
	\label{tab:spatial distribution analysis}
	\begin{tabular}{c c c c c c c c} 
		\bottomrule
		\hline
		Model & Position & Extension ($^\circ$) & Spectral Index  & TS Value & $TS_{\mathrm{ext}}$ \\
		\hline
		Point source & $1.76{^\circ}  \pm {0.06{^\circ}}$,$73.02{^\circ}  \pm {0.06{^\circ}}$ & ... & 2.06 $\pm$ 0.78 & 27.60 & ... \\
		Uniform disk & $1.68{^\circ}  \pm {0.04{^\circ}}$,$73.17{^\circ}  \pm {0.04{^\circ}}$ & $0.41{^\circ} \pm {0.03{^\circ}}$ & 1.61 $\pm$ 0.36 &  44.94 & 28.33  \\
		2D Gaussian & $1.65{^\circ}  \pm {0.05{^\circ}}$,$73.15{^\circ}  \pm {0.05{^\circ}}$ & $0.38{^\circ} \pm {0.10{^\circ}}$ & 1.68 $\pm$ 0.35 &  42.10 & 27.61 \\
		\hline
	\end{tabular}

\end{table*}

\begin{table}
    \centering
	\caption{The Energy Flux of SrcA with Fermi-LAT Data in the 50 GeV$-$1 TeV Energy Band.}
	\label{tab:flux}
        \setlength{\tabcolsep}{7pt}
	\begin{tabular}{lcc c c c c}
		\bottomrule
		\hline
    E & Band & E$^2$ dN/dE & TS Value\\
     (MeV) & (MeV) & (10$^{-12}$ erg cm$^{-2}$ s$^{-1}$) & \\
            \hline
      82377  & 50000$-$135721   & 1.38 $\pm$ 0.50${_{-0.19}^{+0.02}}$ & 17.68 & \\
     223606  & 135721$-$368400  & 1.70 $\pm$ 0.88${_{-0.85}^{+0.84}}$ & 11.41 &\\
     606959  & 368400$-$1000000  & 3.18 $\pm$ 1.92${_{-2.49}^{+1.54}}$ & 12.27 &\\
        \hline
	\end{tabular}
 \flushleft{{\bf Notes.} The energy fluxes of SrcA with 1$\sigma$ statistic error in each energy bin are given, and the second one indicates systematic error.}
\end{table}

\cite{2013ApJ...773...77A} collected a large number of PWNe detected by Fermi-LAT and studied the correlations between different energy bands, establishing new constraints. Using 45 months of Fermi-LAT data, they derived an upper limit of 6.9 $\times$ $10^{-12}$ erg $\mathrm{cm}^{-2}$ $\mathrm{s}^{-1}$ on the flux of CTA 1 between the 10$-$316 GeV band. \cite{2016ApJ...831...19L} performed a detailed analysis of the radiation properties of PSR J0007+7303 during the off-peak and the on-peak phase intervals. However, no point-like or extended $\gamma$-ray emission between 10 and 300 GeV was detected during the off-peak phase of PSR J0007+7303, resulting an upper limit of $6.5 \times 10^{-12} \,\mathrm{erg} \,\mathrm{cm}^{-2} \,\mathrm{s}^{-1}$ instead. According to The Third Fermi Large Area Telescope Catalog of Gamma-ray Pulsars \citep{2023arXiv230711132S}, the spectral energy distribution (SED) of PSR J0007+7303 peaks at 2.44 $\pm$ 0.03 GeV. We assumed that the contribution of pulsar to GeV flux can be described by a PLEC4 model \citep[For details, see][]{2022ApJS..260...53A}. As shown in Figure \ref{fig:sedHigh}, the flux of the pulsar exponentially decays in the GeV band. When selecting the energy range from 50 GeV to 1 TeV, an extended source in the region overlapping observations from other telescopes exhibits GeV $\gamma$-ray emission with a hard spectral index of $\sim$ 1.61. This indicates that the high-energy radiation may originate from the PWN in CTA 1. This suggests that above 50 GeV, the pulsar no longer contributes significantly to the $\gamma$-ray emission. Therefore, to obtain higher photon statistics, the utilization of pulsar gating is unnecessary.

Based on the uniform disk model with an extension of $R_{\mathrm{68}}$ = $0.41{^\circ} \pm {0.03{^\circ}}$, we obtained a global fitted flux of (6.71 $\pm$ 2.60) $\times$ $10^{-12}$ erg $\mathrm{cm}^{-2}$ $\mathrm{s}^{-1}$ with a spectral index of 1.61 $\pm$ 0.36 and a TS value of $\sim$ 44.94 calculated by the likelihood analysis in the energy range of 50 GeV$-$1 TeV. We divided the data from 50 GeV to 1 TeV band into 3 logarithmically equal energy bins and performed the same likelihood analysis for each bin. The results are displayed as blue dots in Figure \ref{fig:sedHigh}, with specific values listed in Table \ref{tab:flux}. In the Fermi High-Latitude Extended Sources Catalog \citep{2018ApJS..237...32A}, a significantly extended ($\sim$ 0.$^\circ$98) $\gamma$-ray source (FHES J0006.7+7314) was reported, which is thought to be associated with the CTA 1. However, there was a mismatch in flux normalization observed between the two spectra. In our work, the derived flux does not exceed the upper limit defined in previous work \citep{2013ApJ...773...77A,2016ApJ...831...19L}.
The obtained hard spectral index value of $\sim$ 1.61 agrees with the typical hard spectral index observed in PWNe by Fermi-LAT \citep{2013ApJ...773...77A}. Furthermore, as shown in Figure \ref{fig:CTA}, the GeV SED could smoothly connect to the TeV one.

\subsection{Systematic Uncertainties Analysis}
\label{systematic uncertainties analysis}

The systematic errors in the Fermi-LAT data are considered from two main origins: imperfect modelling of the Galactic diffuse emission and uncertainties regarding the calibration of the effective area. For the first aspect, we estimated the systematic error by repeated analyses using the Galactic diffuse model with artificially altered normalization of $\pm 6\%$ \citep{2010ApJ...722.1303A,2013ApJS..208...17A}, and for the second aspect, we used the bracketing Aeff method\footnote{\url{https://fermi.gsfc.nasa.gov/ssc/data/analysis/scitools/Aeff_Systematics.html}} \citep{2012ApJS..203....4A}. The specific systematic error values we listed in Table \ref{tab:flux}.

\section{Discussion} \label{Discussion}

\begin{figure*}
 \centering
	\includegraphics[width=1.8\columnwidth]{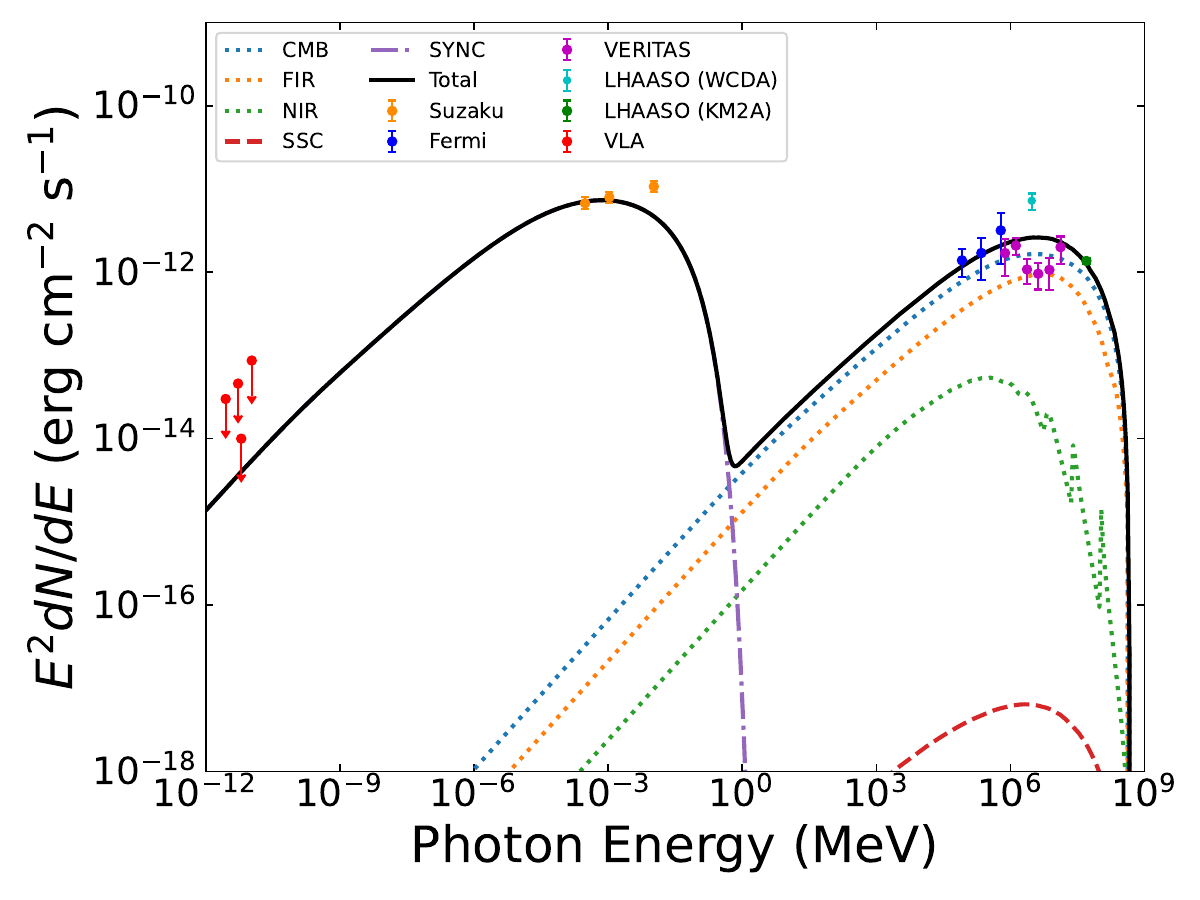}
    \caption{Broad-band SED of the PWN model at t$_{\mathrm{age}}$ = 10 kyr with observed data. The purple dotted line represents the synchrotron radiation, and the red, blue, orange and green dotted lines are for the inverse Compton scatterings off the synchrotron photons, CMB and FIR and NIR background, respectively. The total SED is shown by the black solid line. The radio data \citep{2013ApJ...764...38A,2013ICRC...33.2656G} and the X-ray data \citep{2012MNRAS.426.2283L} are shown in this figure as red and orange dots, respectively. The fluxes in GeV $\gamma$-rays with the statistic error (this work) and in TeV $\gamma$-rays with $VERITAS$ \citep{2013ApJ...764...38A} and LHAASO \citep{2023arXiv230517030C} are also shown in this figure.}
    \label{fig:CTA}
\end{figure*}

\begin{figure*}
 \centering
	\includegraphics[width=1.8\columnwidth]{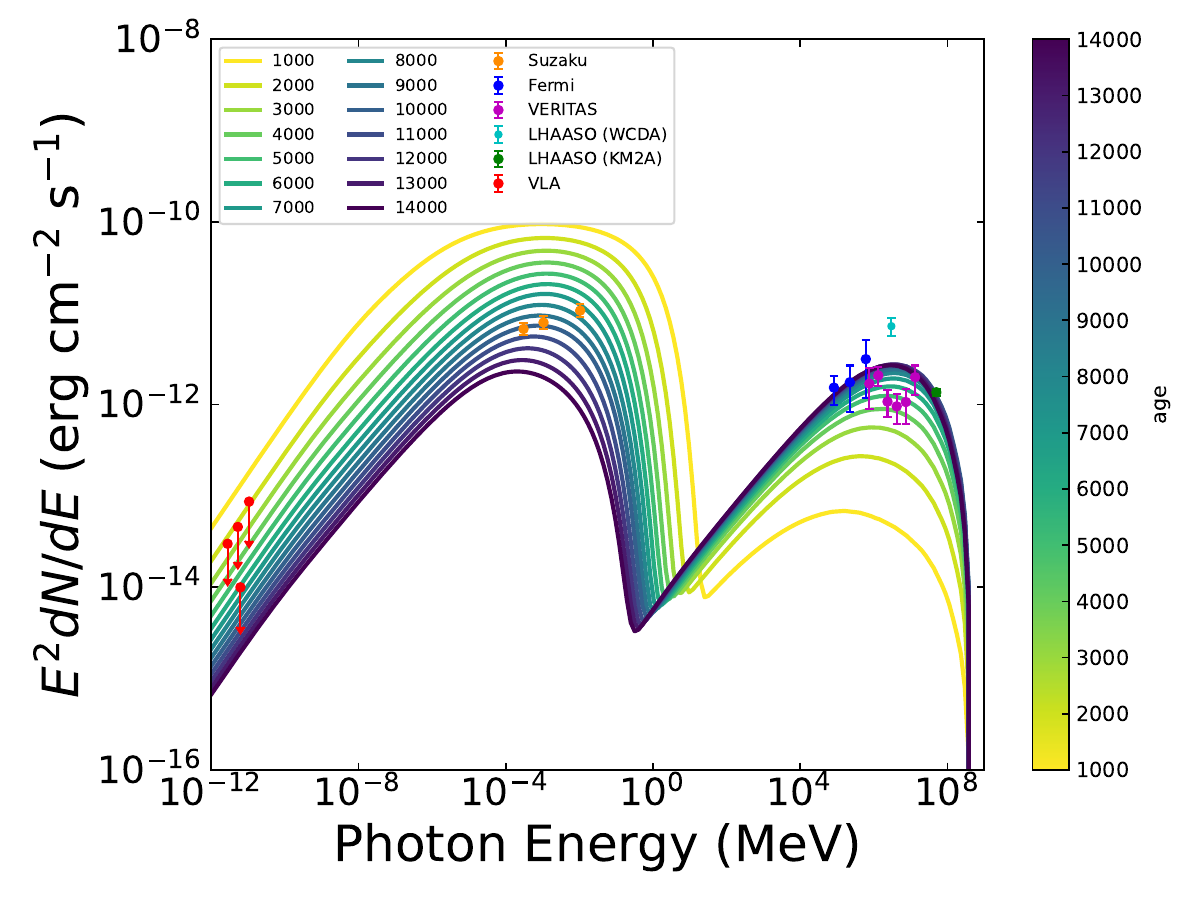}
    \caption{Time evolution of the SED from the PWN model. The color bars and lines using a continuous color scale show the SED of nebula at different ages from 1 to 14 kyr. The representation of the other data points is consistent with that of Fig. \ref{fig:CTA}.}
    \label{fig:CTAnianling}
\end{figure*}

\begin{table}
    \centering
	\caption{Fitting Parameters of the PWN model}
	\label{tab:pwnparameters}
        \setlength{\tabcolsep}{12pt}
	\begin{tabular}{lcc c c}
		\bottomrule
		\hline
    Parameters &  \\
            \hline
      Age (t$_{\mathrm{age}}$,kyr)  & 10   \\
      Initial spin-down luminosity ($L_0$,$\,\mathrm{erg}\,\mathrm{s}^{-1}$) & $2.78\times10^{36}$ \\
      Current spin-down luminosity ($L_t$,$\,\mathrm{erg}\,\mathrm{s}^{-1}$) & $4.48\times10^{35}$ \\
      Initial spin-down timescale ($\tau_0$,yr)& 3900 \\
      Particle index ($\alpha_1$) & 1.8 \\
      Particle index ($\alpha_2$) & 2.0 \\
      Break Lorentz factor ($\gamma_b$) & 8 $\times$ 10$^4$ \\
      Magnetic energy ratio ($\eta$) & 0.4 \\
      Magnetic field strength (B,$\mu$G) & 3.25 \\

        \hline
	\end{tabular}
\end{table}

The hadronic origin of VER J0006+729 has been extensively investigated. According to the findings of \cite{2016MNRAS.459.3868M}, the mass of the molecular clouds is insufficient to explain the observed TeV emission. Additionally, \cite{2013ApJ...764...38A} found that the much smaller extent of VER J0006+729 compared to the supernova remnant (SNR) contradicts the SNR shell hypothesis. Therefore, we favour the leptonic PWN origin scenario for VER J0006+729.

The SED of PWNe exhibit a bimodal structure, with the low-energy component generated through synchrotron from radio to X-ray band, while the high-energy component in the $\gamma$-ray band is produced by inverse Compton scattering off the soft photon field. \cite{2013ApJ...764...38A} used the 1.4 GHz image from \cite{1997A&A...324.1152P} to estimate the radio flux within a 20 arcmin radius around the pulsar, and here we adopt this flux as an upper limit for the radio emission. \cite{2013ICRC...33.2656G} also presented new high angular resolution and high sensitivity radio observations toward PSR J0007+7303 at 1.5 GHz with the Jansky Very Large Array, and set an upper limit considering a size for the nebula of 20 arcmin in radius. In the X-ray band, \cite{2012MNRAS.426.2283L} detected a $\sim$10 arcmin extended feature with $Suzaku$ that may correspond to the bow shock of the nebula. In the TeV $\gamma$-ray band, the TeV source VER J0006+729 associated with CTA 1 was detected by $VERITAS$ \citep{2013ApJ...764...38A}, which has a photon spectrum described by a power-law spectrum $dN/dE = N_0\left( E/3 \, TeV \right) ^{-\varGamma}$ with a differential spectral index of $\varGamma = 2.2\pm 0.2_{stat}\pm 0.3_{sys}$ and the normalization $N_0 = \left( 9.1\pm 1.3_{stat}\pm 1.7_{sys} \right) \times 10^{-14}cm^{-2}s^{-1}TeV^{-1}$. CTA 1 is also thought to be associated with a UHE source 1LHAASO J0007+7303u catalogued in the first LHAASO catalog \citep{2023arXiv230517030C}, with a flux of $(5.01\pm 1.11)\times 10^{-13} \, TeV^{-1}cm^{-2}s^{-1}$ detected by WCDA at 3 TeV and a flux of $(3.41\pm 0.27)\times 10^{-16} \, TeV^{-1}cm^{-2}s^{-1}$ detected by KM2A at 50 TeV. The aforementioned observational data, as well as the Fermi-LAT data analyzed in this work, are plotted in Fig. \ref{fig:CTA}.

We considerd a time-dependent one-zone model for the radiative evolution of a PWN \citep[For details, see][]{2014JHEAp...1...31T,2018A&A...612A...2H}. The spin-down power of an energetic pulsar is continuously transferred to high-energy particles and magnetic field in the PWN, and the spin-down power L(t) evolves in time as \citep[][]{2006ARA&A..44...17G} 
\begin{equation}
    L(t) = L_0 {( 1+\frac{t}{\tau_0} )}^{-\frac{n+1}{n-1}},
\end{equation}
where $L_0(t)$ is the initial luminosity, the breaking index n is assumed to be 3.0 here. And the initial spin-down time-scale $\tau_0$ of the pulsar is \citep[][]{2006ARA&A..44...17G}
\begin{equation}
    \tau_0 = \frac{2\tau_c}{n-1}-t_{\mathrm{age}},
\end{equation}
where $\tau_c = {P}/{2 \dot{P}}$ is the characteristic age can be derived from the period and the period derivative.

The particle distribution $N\left( \gamma ,t \right)$ follows the equilibrium equation \citep[e.g.][]{2012MNRAS.427..415M}
\begin{equation}
   \frac{\partial N(\gamma,t)}{\partial t}= - \frac{\partial }{\partial {\gamma}}[\dot{\gamma}(\gamma,t)N(\gamma,t)] - \frac{N(\gamma,t)}{\tau(\gamma,t)} + Q(\gamma,t).
\end{equation}
We assumed that the injection rate of particles $Q(\gamma,t)$ follow a broken power-law distribution, i.e.,
\begin{equation}
   Q(\gamma,t)= Q_0(t)
   \begin{cases}
   (\frac{\gamma}{\gamma_\mathrm{b}})^{\mathrm{-\alpha_1}} \quad if \gamma \le \gamma_\mathrm{b} \\\\
   (\frac{\gamma}{\gamma_\mathrm{b}})^{\mathrm{-\alpha_\mathrm{2}}} \quad if \gamma_\mathrm{b} < \gamma \le \gamma_\mathrm{max}
   \end{cases} ,
\end{equation}
in which ${\gamma_b}$ is the break Lorentz factor, and ${\alpha_1}$ and ${\alpha_2}$ are the spectral indices.

The expansion behaviour of the PWN in the SNR environment is determined by the age of the system t, the initial spin-down time-scale $\tau_0$, and the reverse-shock interaction time $t_{rs}$, which as follows \citep{2012arXiv1202.1455M,2018A&A...612A...2H}:
\begin{equation}
R(t)\quad\propto\quad\begin{cases}t^{6/5} & \text{for }t\leqslant\tau_0\\ t & \text{for }\tau_0<t\leqslant t_{\text{rs}}\\ t^{3/10} & \text{for }t>t_{\text{rs}}.\end{cases}
\end{equation}

The magnetic field strength evolves with time can be calculated by \citep{2010ApJ...715.1248T,2012MNRAS.427..415M}
\begin{equation}
    {B(t) = \sqrt{\frac{3(n-1) \eta L_{\mathrm{0}} \tau_{\mathrm{0}}} {R_{\mathrm{PWN}}^{3}(t)} \left[ 1-\left(1+\frac{t}{\tau_{\mathrm{0}}}\right)^{-\frac{2}{n-1}}\right] } .}
\end{equation}
The young pulsar PSR J0007+7303 in CTA 1 has a period of $\sim$316 ms, a spin-down power of $\sim$$4.5 \times 10^{35} \mathrm{erg}\, \mathrm{s}^{-1}$ and a characteristic age of $\tau_c$ = 13 kyr \citep{2012ApJ...744..146A}, which is energetic to power the nebula \citep{2004ApJ...601.1045S}. We adopt a distance of 1.4 kpc in this paper \citep{1993AJ....105.1060P}. The interstellar radiation field including CMB, the near-infrared (NIR) background with $T_{\mathrm{NIR}}$ = 25.0 K and $U_{\mathrm{NIR}}$ = 0.3 eV cm$^{-3}$ and far-infrared (FIR) background with $T_{\mathrm{FIR}}$ = 3000 K and $U_{\mathrm{FIR}}$ = 0.6 eV cm$^{-3}$ were calculated using GALPROP code \citep{2006ApJ...648L..29P} by \cite{2018A&A...609A.110Z}. We assumed that the nebula has an age of 10 kyr, and that the corresponding $\tau_0$ and $L_0$ are 3900 yr and $2.78\times10^{36}\,\mathrm{erg} \,\mathrm{s}^{-1}$. For the particle spectrum, we assumed a broken power-law with $\alpha_1$ = 1.8, $\alpha_2$ = 2.0 and $\gamma_b$ = 8 $\times$ 10$^4$, and the resulting SED at t$_{\mathrm{age}}$ is shown in Fig. \ref{fig:CTA}. With the magnetic energy ratio $\eta$ = 0.4, the magnetic field strength in the nebula is 3.25 $\mu$G.

As the PWNe evolve continuously inside the SNR, the expansion of the PWNe results in a gradual decrease in magnetic field strength \citep{2014JHEAp...1...31T}. In the model proposed by \cite{2009ApJ...703.2051G}, as the PWNe evolve to about 10 kyr, the magnetic field strength decreases to a few $\mu$G. In this work, we obtained a relatively low magnetic field strength of 3.25 $\mu$G. Similar low values have also been adopted for several PWNe, such as HESS J1826$-$130 \citep{2022ApJ...930..148B}, LHAASO J1908+0621 \citep{2022ApJ...934..118D,2021ApJ...913L..33L,2021MNRAS.505.2309C}, LHAASO J2226+6057 \citep{2023MNRAS.520.5858J,2022A&A...668A..23D}, and HESS J1303$-$631 \citep{2012A&A...548A..46H}.

As shown in Fig. \ref{fig:CTA}, with these reasonable parameters, the observed SED can be well reproduced. In addition, we present in Fig. \ref{fig:CTAnianling} the evolution of the SED at different
ages from 1 to 14 kyr, which is able to fit the observation data points well when the nebula age is 10 kyr. Finally, the fitting parameters we used for the model are summarized in Table \ref{tab:pwnparameters}.

It is well known that PWNe are a prominent population of TeV sources in the Milky Way \footnote{\url{http://tevcat.uchicago.edu/}} as observed by various instruments in recent years \citep{2015ICRC...34..771P,2018A&A...612A...2H,2020ApJ...905...76A}. In particular, the detection of photons with energies up to $\sim$ 1 PeV from the Crab Nebula implies the presence of PeV electron accelerators (Pevatrons) \citep{2021Sci...373..425L}. Observations from LHAASO have revealed numerous hundred TeV sources in the Milky Way \citep{2021Natur.594...33C,2023arXiv230517030C}, a significant fraction of which can be attributed to PWNe \citep{2023MNRAS.519.1881W,2023PASP..135g4503T,2023RAA....23j5003X} and are plausible candidates for PeVatrons \citep{2022ApJ...930L...2D}. The detection of UHE $\gamma$-rays from CTA 1, suggests its potential as a PeVatron. Future observations in the UHE band can provide further insights into the energy distribution of particles within CTA 1.

\section{Summary} \label{Summary}

In this paper, we analyzed the GeV $\gamma$-ray emission toward CTA 1 using about 15 yr of Fermi-LAT data. Under the hypothesis of a uniform disk model with an extension of $R_{\mathrm{68}}$ = $0.41{^\circ} \pm {0.03{^\circ}}$, we obtained a global flux of (6.71 $\pm$ 2.60) $\times$ $10^{-12}$ erg $\mathrm{cm}^{-2}$ $\mathrm{s}^{-1}$ with a spectral index of 1.61 $\pm$ 0.36 and a TS value of $\sim$ 44.94 in the 50 GeV$-$1 TeV energy band. The GeV emission could potentially be from the PWN in the CTA 1, wherein the source is powered by PSR J0007+7303 and emits $\gamma$-rays through inverse Compton scattering. The SED could be reasonably reproduced by a time-dependent one-zone model with a broken power-law spectrum. CTA 1 has also been detected by $VERITAS$ and LHAASO instruments. The UHE $\gamma$-ray emission detected by LHAASO implies the potential for particle acceleration to the PeV range. Further in-depth observations and analyses are needed to clarify the maximum energy of the particles in the PWN.

\section*{Acknowledgements}

We thank the public data provided by Fermi-LAT. This research is supported by NSFC through Grants 12063004 and 12393852, as well as grants from the Yunnan Provincial Government (YNWR-QNBJ-2018-049), Yunnan Fundamental Research Projects (grant No. 202201BF070001-020), and the Program of Yunnan University (KC-22221102).

\section*{DATA AVAILABILITY}
The data produced in this paper will be shared on reasonable request to the corresponding author.




\bibliographystyle{mnras}
\bibliography{MN-23-4553-MJ} 





\bsp	
\label{lastpage}
\end{document}